\documentclass[twocolumn,secnumarabic,amssymb, nobibnotes, aps, prd]{revtex4-2}

\usepackage{amsmath}
\usepackage{amsfonts}
\usepackage{amssymb}

\usepackage{psfrag,graphicx}
\usepackage{float}
\usepackage[T1]{fontenc}
\usepackage{xcolor}
\usepackage{lmodern}
\usepackage{listings}
\usepackage{subfigure}
\usepackage[section]{placeins}
\usepackage{psfrag}
\usepackage{rotating}
\usepackage{multirow}
\usepackage{colortbl}
\allowdisplaybreaks
\usepackage{parskip}
\usepackage[T1]{fontenc}

\begin{document} 
\title{Unidirectional segregation of Bright-Bright soliton through a $\mathcal{PT}$-symmetric potential}%

\author{M. O. D.~Alotaibi}
\affiliation{Department of Physics, Kuwait University, P.O. Box 5969 Safat, 13060, Kuwait}
\email[majed.alotaibi@ku.edu.kw]
\date
\begin{abstract}
We study the dynamics of two-component vector solitons, namely, bright-bright (BB) solitons interacting with parity-time-($\mathcal{PT}$) symmetric potentials. We employ direct numerical simulations to demonstrate the unidirectional segregation of the BB soliton. Using a modified perturbed dynamical variational Lagrangian approximation, we develop an analytical model to verify the results obtained from numerical calculations. Simplified variational equations of motion suggest that the splitting of BB solitons can be explained by considering the effective force between the two components.  
\end{abstract}
\maketitle
 

\section{Introduction}
\label{sec:Introduction}

Unidirectional flow is an important topic in physical research from fundamental and applied perspectives. It has been used in many fields of applied physics, such as electromagnetic waves~\cite{2,3}, phonon waves~\cite{5,6,7}, and metamaterials~\cite{8,9}. In nonlinear wave theory, controlling the flow direction of solitons has profound applications owing to the appealing features that solitons provide for optical data transfer and processing~\cite{book1,book2,book3,book4}. The scattering of bright solitons described by the nonlinear Schr\"{o}dinger equations with a reflectionless potential has been extensively studied ~\cite{21,22,12a}. In addition, the unidirectional flow of one-component bright solitons through a specific combination of asymmetric potential wells has been demonstrated in~\cite{10}. Moreover, such a flow of solitons was found to occur in solitons scattered through $\mathcal{PT}$-symmetric potentials~\cite{11}. 

$\mathcal{PT}$-symmetric potentials~\cite{12b} in quantum mechanics are complex potentials that exhibit a purely real spectrum of energies~\cite{13,14,15,16,17}. For example, a one-dimensional Hamiltonian is $\mathcal{PT}$-symmetric when the corresponding potential fulfils the condition  $V(x)=V^{*}(-x)$, where $x$ is the spatial coordinate and the asterisk denotes a complex conjugation. In this case, the real part of the potential must be an even function of position $x$, whereas the imaginary part must be an odd function.  

For a two-component bright-bright (BB) soliton interacting with asymmetric double wells or barriers, the flow direction of each component could be controlled separately~\cite{12}. In this case, we may obtain a unidirectional flow for both components or split the BB soliton into its constituents as desired, allowing more flexibility for manipulating such systems. Therefore, it is important to search for different schemes by which it is possible to achieve unidirectional segregation/splitting of vector solitons, such as BB solitons. To the best of our knowledge, no previously reported study has treated the unidirectional segregation of BB solitons using $\mathcal{PT}$-symmetric potential, which is the subject of our present study.

Here, we performed numerical simulations and theoretical analysis to study BB soliton scattering through a $\mathcal{PT}$-symmetric potential. We showed from numerical results that unidirectional segregation can be obtained using a reflection-less potential type, so-called the Rosen-Morse (RM) potential. We also provided analytical and numerical proves of the unidirectional segregation with the delta function in the real and imaginary parts of the $\mathcal{PT}$-symmetric potential. Furthermore, we performed analytical calculations using a modified perturbed dynamical variational Lagrangian approximation method. The outcome of the variational calculations provides insight into the physics behind the splitting of the two components and determines the upper limit of the coupling strength constant for breaking the BB soliton into its components. 

The remainder of this paper is organized as follows. In Sec.~\ref{sec:Problem_Formulation} we introduce the proposed theoretical model. In Sec.~\ref{sec:Numerical_Results}, we numerically prove the BB soliton unidirectional segregation when passing through a $\mathcal{PT}$-symmetric potential, where the barrier has been modelled using the RM potentials. Section~\ref{sec:Variational_Approach_versus_Numerical_Computation} 
presents a comparison between the variational approach and numerical simulation using delta function potentials for the real and imaginary parts of the $\mathcal{PT}$-symmetric potential. Finally, in Sec.~\ref{sec:Conclusions} we summarize our findings.





\section{Problem Formulation}
\label{sec:Problem_Formulation}
The normalized nonlinear Schr\"{o}dinger equation for bright-bright vector solitons with a $\mathcal{PT}$-symmetric potential is given by
\begin{align}
	\label{eq:coupled_NLSE}
&i \frac{\partial}{\partial t} u  +\frac{1}{2} \frac{\partial^2}{\partial x^2} u  + \left[|u|^2    + g|v|^2 \right] u +U(x) u =0, \\ \nonumber
&i \frac{\partial}{\partial t} v  +\frac{1}{2} \frac{\partial^2}{\partial x^2} v   + \left[g |u|^2  + |v|^2 \right] v  +U(x) v = 0, 
\end{align}

where $u\equiv u(x,t)$ and $v\equiv v(x,t)$  are the wave functions for the bright-bright vector soliton components. The coupling strength between the two components is controlled by $g$, that is, for $g=0$, the system is completely decoupled and two bright solitons are obtained. The potential in the above equations takes the form
\begin{align}
	\label{eq:pt_potential}
 U(x) = V\left(x\right) + i W\left(x\right),  
\end{align}
for both components, where $V(x)$ and $W(x)$ are the even and odd functions, respectively. 

In Sec.~\ref{sec:Numerical_Results}, we begin our analysis by numerically proving the unidirectional segregation of the BB soliton, where we set  
\begin{align}
	\label{eq:real_imaginary_potential_1}
 V(x) = V_0 \; \mathrm{sech}^2\left(\alpha x \right), \; W(x) =\beta \; W_0 \; x \; \mathrm{sech}^2\left(\alpha x \right).
\end{align}
The constants $V_0$ and $W_0$ are real-valued constants that correspond to the depth or amplitude of the real and imaginary parts of the potential, respectively. The real part in Eq.~\eqref{eq:pt_potential} represents a class of reflectionless potentials known as the Rosen-Morse potential, where $\alpha$, the inverse width, is usually equal to $\sqrt{|V_0|}$ to maintain the reflectionless property. The constant, $\beta$, in Eq.~\eqref{eq:real_imaginary_potential_1}, and Eq.~\eqref{eq:real_imaginary_potential_2} reflects the potential around $x=0$ by setting $\beta=\pm 1$. 

In Sec.~\ref{sec:Variational_Approach_versus_Numerical_Computation}, we chose
\begin{align}
	\label{eq:real_imaginary_potential_2}
 V(x) = V_0 \delta\left(x\right), \; W(x) =\beta \; W_0 \left[\delta\left(x-L\right) - \delta\left(x+L\right)\right], 
\end{align}
to obtain analytical expressions using variational calculations. The $\mathcal{PT}$-symmetry requirement is satisfied for both the potentials, as shown in Eq.~\eqref{eq:real_imaginary_potential_1}, and Eq.~\eqref{eq:real_imaginary_potential_2}~\cite{11}.

To study the unidirectional segregation/splitting of the BB vector soliton, the potential is fixed at the center, BB vector soliton is launched from both sides, and the scattered region is thus observed. Our analysis follows an equivalent plot, where we fix the BB vector soliton launching point and rotate the $\mathcal{PT}$-symmetric potential around $x=0$ by switching $\beta$ from $+1$ to $-1$ (see Fig. ~\ref{fig:FRHPRA:den_poten_-1/fig3} and Fig. ~\ref{fig:FRHPRA:den_poten_1/fig3}).  

\FloatBarrier


\section{Numerical Results Using RM Potentials}
\label{sec:Numerical_Results}
We numerically studied the interaction between the BB soliton and the potential described by Eq.~\eqref{eq:real_imaginary_potential_1}. It has been shown that solitons scattered by this potential display a sharp transition in transport coefficients at a specific critical incident center-of-mass speed of the soliton~\cite{11,19,20}. Therefore, it is natural to start the analysis using this potential to prove the unidirectional segregation phenomenon of the BB soliton. However, the product of hyperbolic secants with different widths cannot be integrated into an analytical form. Therefore, we selected different potential functions in Sec. ~\ref{sec:Variational_Approach_versus_Numerical_Computation} for the variational and numerical calculations. In the absence of the potential term, Eq.~\eqref{eq:coupled_NLSE} has an analytical solution of the following form
  
\begin{align}
	\label{eq:ansatz_num1}
 u\left(x,0\right) &= A \mathrm{sech} \left[\frac{x-\xi_{1}}{a}\right] e^{-i v_{1} x}, \\ \nonumber 
 v\left(x,0\right) &= A \mathrm{sech} \left[\frac{x-\xi_{2}}{a}\right] e^{-i v_{2} x},
\end{align}
where $\xi_{1,2}$ and $v_{1,2}$ represent the soliton location and velocity, respectively. The soliton amplitude is set to $A=\frac{1}{\sqrt{2}}$ and width $a=\sqrt{2}$. The initial positions of the two solitons are chosen far from the potential location, such that there is no interaction between them at $t=0$. In our analysis, we fixed the launching point of the vector soliton and rotated the potential, Eq.~\eqref{eq:real_imaginary_potential_1}, around $x = 0$ such that when $\beta = 1$ ($\beta = -1$) it is equivalent to a BB vector soliton coming from the left (right). We set the BB soliton in motion with the initial center-of-mass velocity $v_{1}=v_{2}$ toward the potential region. The results range from total reflection $\left(\textit{R}\right)$, transmission $\left(\textit{T}\right)$, trapping $\left(\textit{L}\right)$, or a combination of these states, depending on the interaction between the BB vector soliton and the potential. The reflectance, transmittance, and trapping coefficients are defined as follows    

\begin{align}
	\label{eq:RLT}
R_{1,2} =& \frac{1}{N}\int_{-\infty}^{-\delta} |\psi_{1,2}\left(x,t\right)|^2 dx, \\ \nonumber
L_{1,2} =& \frac{1}{N}\int_{-\delta}^{\delta}  |\psi_{1,2}\left(x,t\right)|^2 dx, \\ \nonumber
T_{1,2} =& \frac{1}{N}\int_{\delta}^{\infty}   |\psi_{1,2}\left(x,t\right)|^2 dx,
\end{align}
where $\psi_{1,2}\left(x,t\right)$ represents the first $u(x,t)$ and second $v(x,t)$ components. The three coefficients must satisfy the conservation law, $R + T + L = 1 $. The constant $\delta$ represents the position of the measurement of reflectance or transmission, set at a value slightly greater than the position of the potential boundary, and $N = \int_{-\infty}^{\infty} \left(|u \left(x,t\right)|^2 + |v\left(x,t\right)|^2 \right) dx$ is the normalization of the BB vector soliton.

\begin{figure}[!h]
	\centering
\includegraphics[width=0.45\columnwidth]{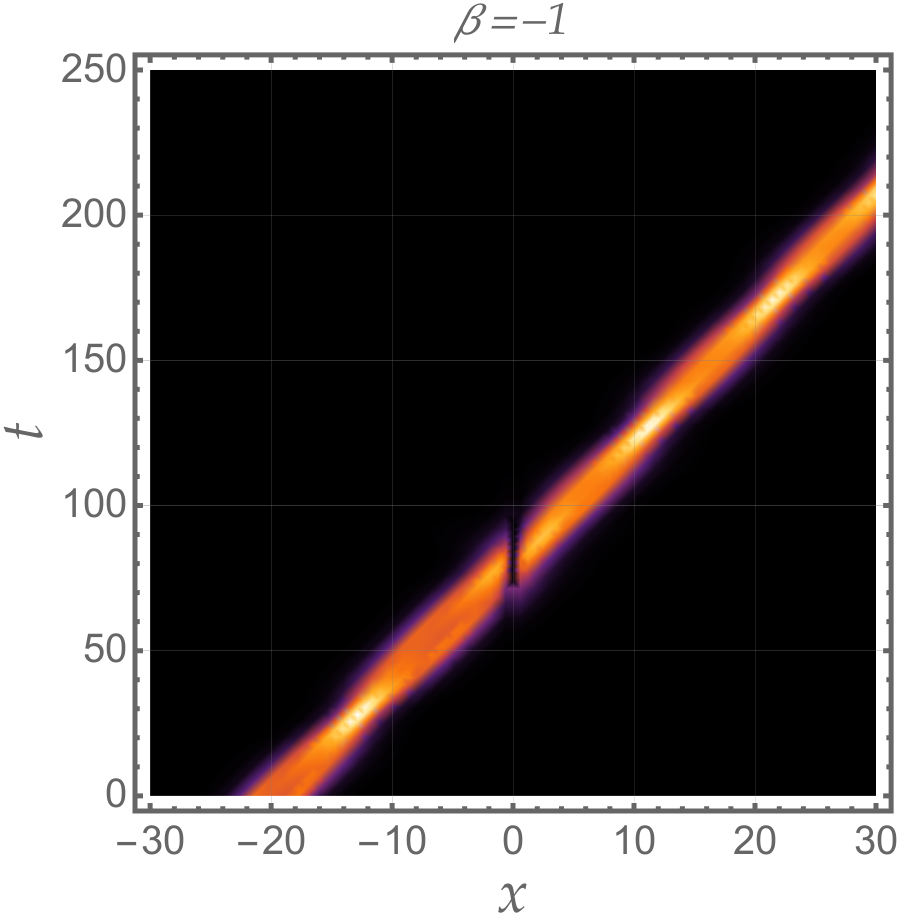}
\includegraphics[width=0.5\columnwidth]{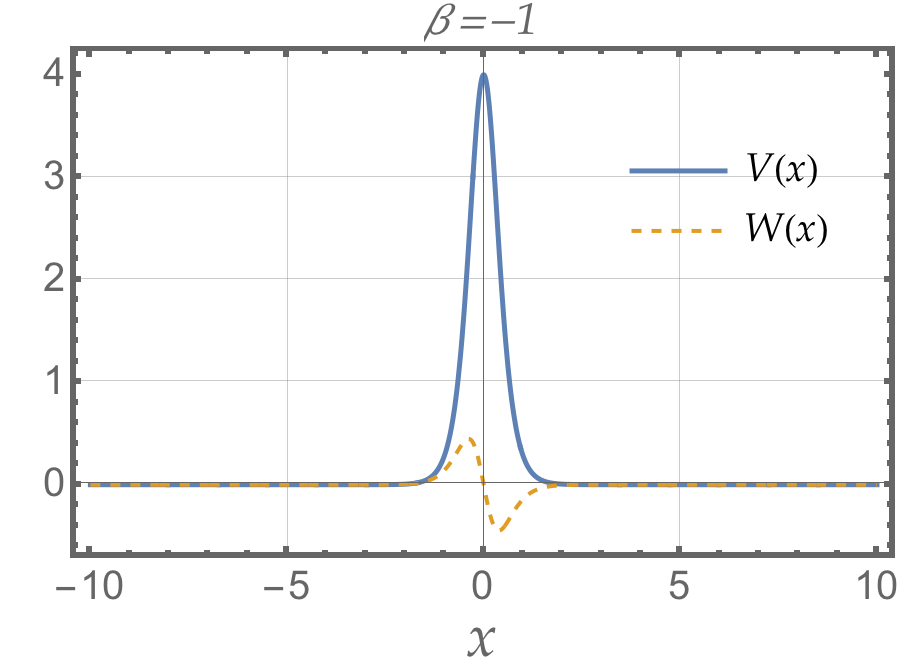}
	\caption{Density plot of a two-component BB soliton (Left panel) interacting with a $\mathcal{PT}$-symmetric potential (right panel). The BB soliton parameters are: $\xi_{1} = -18,\xi_{2} = -21, v_{1}=v_{2}=v_{0}= 0.24$. The potential parameters are: $V_{0}=4, W_{0}=2$. Here we set $\beta=-1$ in Eq.~\eqref{eq:real_imaginary_potential_1}.}
	\label{fig:FRHPRA:den_poten_-1/fig3}
\end{figure}

In Fig.~\ref{fig:FRHPRA:den_poten_-1/fig3} we launched the BB vector soliton with initial positions $\xi_{1}=-18$ and $\xi_{2}=-21$, where the initial velocity was equal to $v_{1}=v_{2}=v_{0}=0.24$. The two components undergo internal oscillation, where we set the interaction coupling $g=0.05$. The potential barrier parameters are $V_{0} = 4, W_{0} = 2$ with $\beta = -1$. In this case, the BB soliton encounters the gain term caused by the imaginary part of the $\mathcal{PT}$-symmetric potential. This results in an increase in its velocity until both components are transmitted over the potential. As shown in Fig.~\ref{fig:FRHPRA:figures/TRL_num_1_beta_minus_1}, this is indeed the case for a wide range of velocities.

\begin{figure}[!h]    
	\centering
\includegraphics[width=0.45\columnwidth]{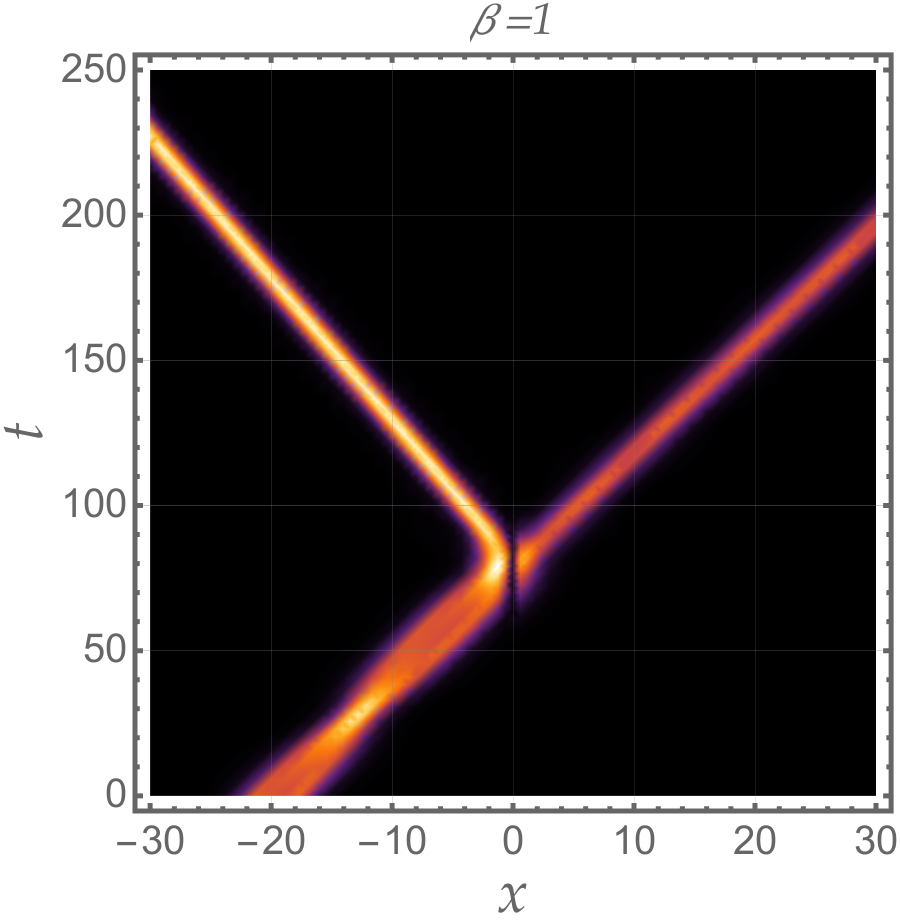}
\includegraphics[width=0.5\columnwidth]{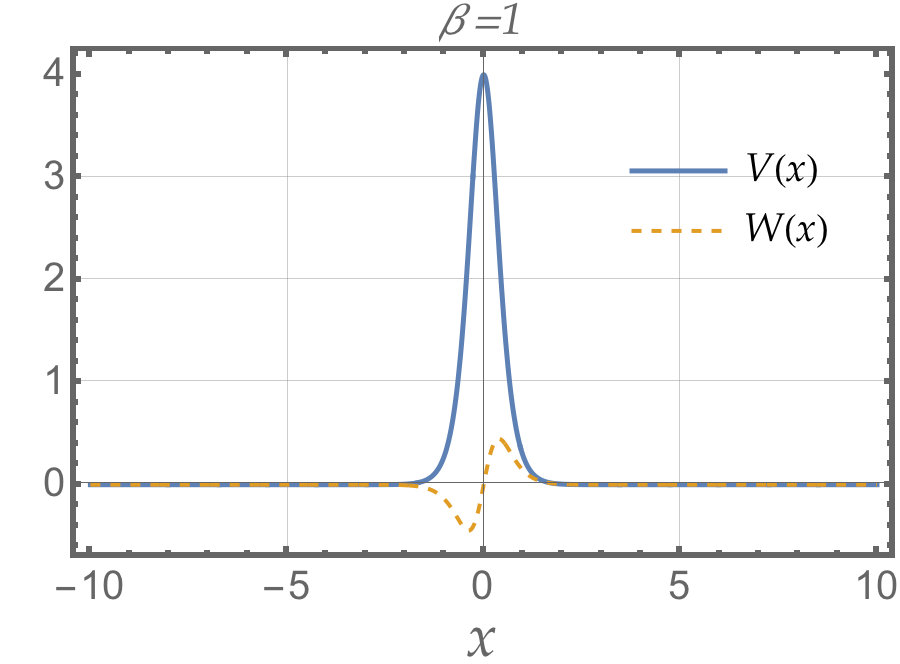}
	\caption{Density plot of two-component BB soliton (Left panel) scattering by a $\mathcal{PT}$-symmetric potential (right panel). We obtain unidierctional segregation using same parameters in Fig.~\ref{fig:FRHPRA:den_poten_-1/fig3} but with $\beta=1$.}
	\label{fig:FRHPRA:den_poten_1/fig3}
\end{figure}

In Fig.~\ref{fig:FRHPRA:den_poten_1/fig3}, we rotate the potential around $x=0$ by setting $\beta = 1$. In this case, the BB soliton first encounters the damping term, which reduces its velocity. Fig.~\ref{fig:FRHPRA:figures/TRL_num_1_beta_pos_1} shows unidirectional propagation may be obtained for both components when $v_{0}<0.17$. In addition, for a wide range of velocities, $0.17 < v_{0} < 0.25$, the unidirectional segregation/splitting of the two components was obtained. For $v_{0}>0.25$, the damping effect caused by the imaginary part of the $\mathcal{PT}$-symmetric potential is not large enough to prevent the two components from being transmitted over the potential. 


 \begin{figure}[!h]
	\centering
\includegraphics[width=0.8\columnwidth]{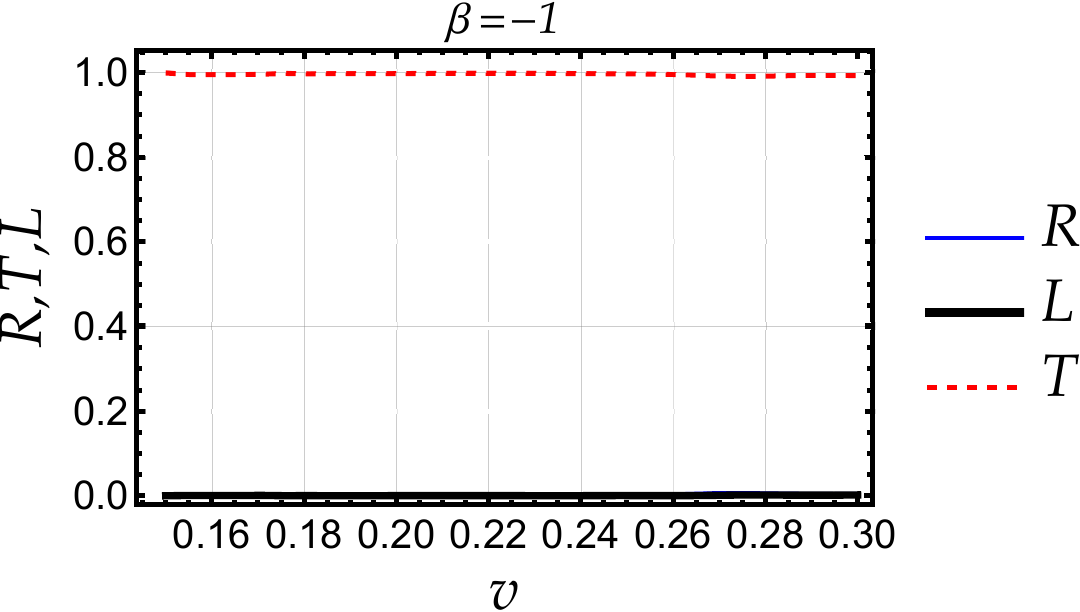}
\includegraphics[width=0.8\columnwidth]{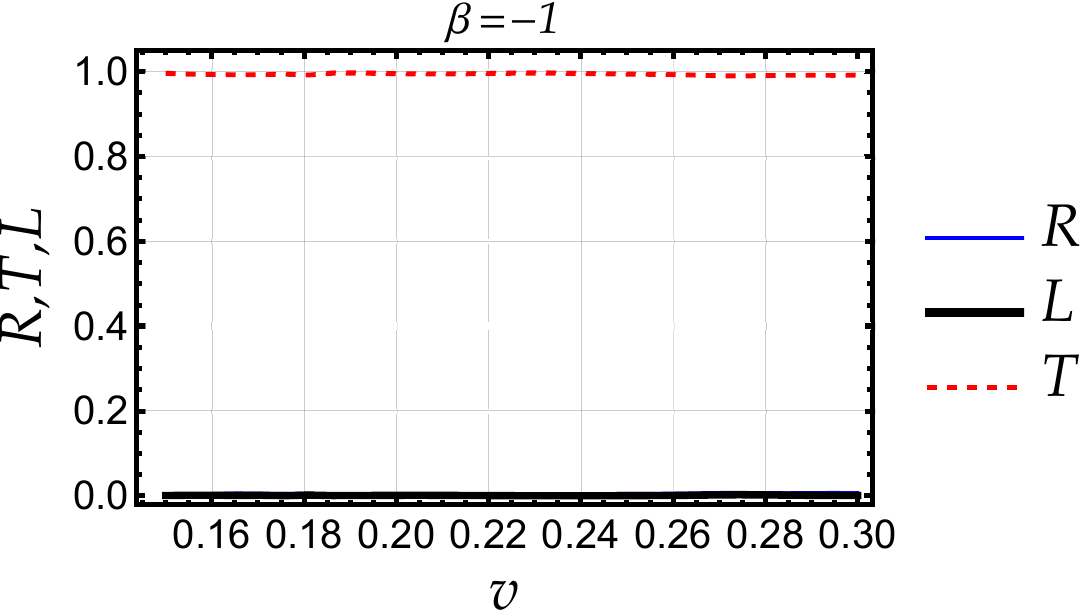}
	\caption{Transport coefficients obtained from the numerical solution of the coupled NLSEs, Eq.~\eqref{eq:coupled_NLSE}, using the potential, Eq.~\eqref{eq:real_imaginary_potential_1} with $\beta=-1$. The up (down) panel represents the first (second) component $u(x,t)$ ($v(x,t)$). For the velocity range, both components transmitted over the potential.}
	\label{fig:FRHPRA:figures/TRL_num_1_beta_minus_1}  
\end{figure}

\begin{figure}[!h]
	\centering
\includegraphics[width=0.8\columnwidth]{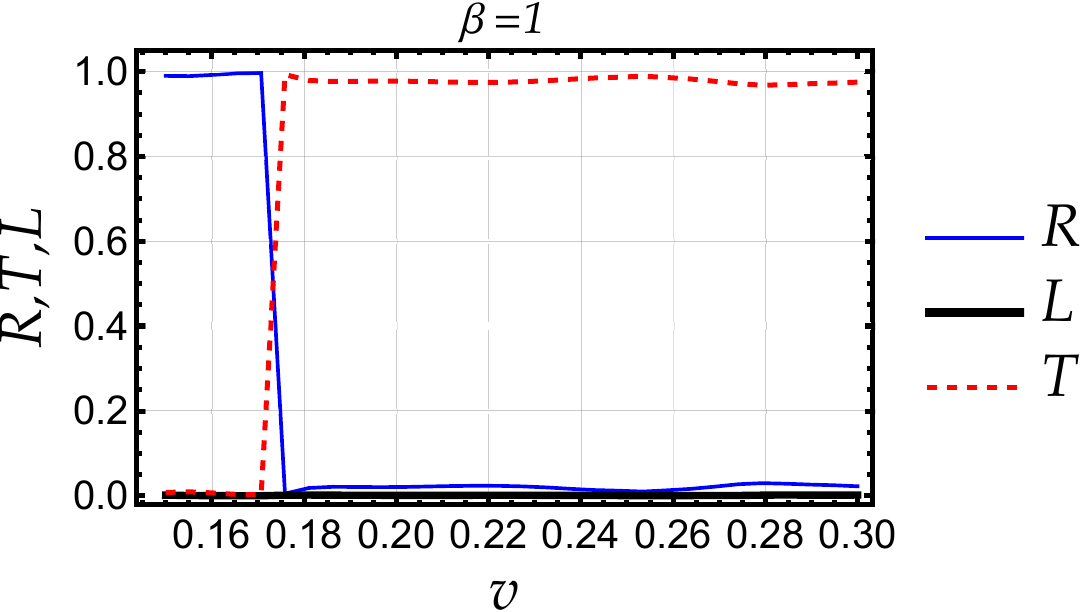}
\includegraphics[width=0.8\columnwidth]{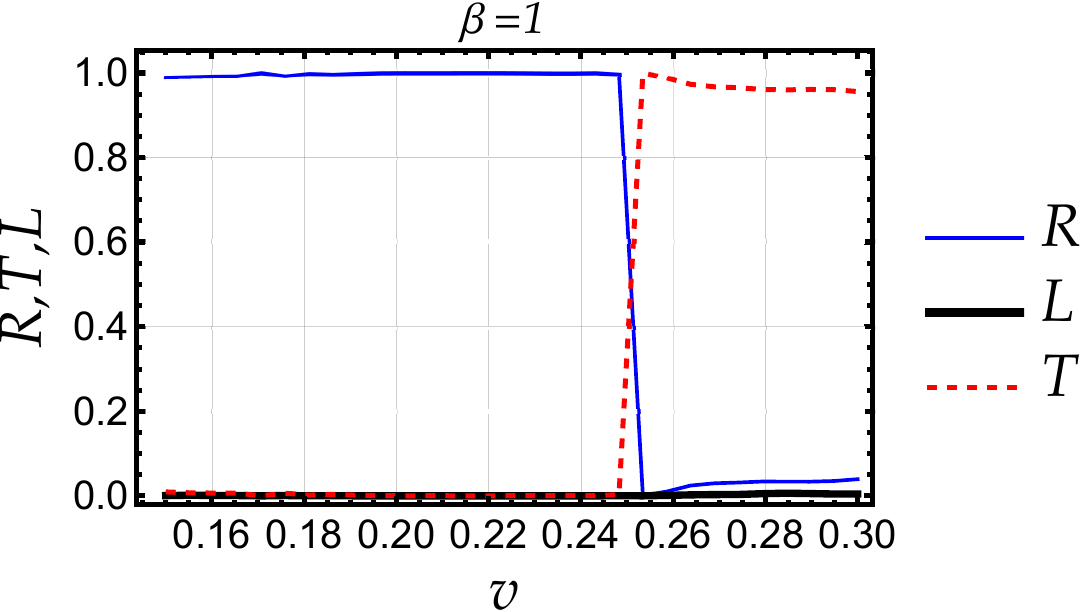}
	\caption{Transport coefficients obtained from the numerical solution of the coupled NLSEs, Eq.~\eqref{eq:coupled_NLSE}, using the potential, Eq.~\eqref{eq:real_imaginary_potential_1} with $\beta=1$. The up (down) panel represents the first (second) component $u(x,t)$ ($v(x,t)$). Depending on the incident velocity, we have a total reflection of both components, unidirectional segregation, or total transmission.}
	\label{fig:FRHPRA:figures/TRL_num_1_beta_pos_1}
\end{figure}

\FloatBarrier


\section{Variational Approach versus Numerical Computation Using Delta Function Potentials}
\label{sec:Variational_Approach_versus_Numerical_Computation}

In this section, we study the interaction between the vector soliton and $\mathcal{PT}$-symmetric potential. The potential term takes the form shown in Eq.~\eqref{eq:real_imaginary_potential_2}. The real and imaginary parts in the potential term, in addition to the assumption $|V_{0}|, |W_{0}| \ll 1$, allow us to treat the potential as a small perturbation effect. Therefore, we can use a modified perturbed dynamic variational Lagrangian approximation. We recast Eq.~\eqref{eq:coupled_NLSE} as follows 
\begin{align}
	\label{eq:coupled_NLSE_new}
&i \frac{\partial}{\partial t} u  +\frac{1}{2} \frac{\partial^2}{\partial x^2} u  + \left[|u|^2    + g|v|^2 \right] u =\epsilon R_{u}, 
\\ \nonumber
&i \frac{\partial}{\partial t} v  +\frac{1}{2} \frac{\partial^2}{\partial x^2} v  + \left[g |u|^2  + |v|^2 \right] v =\epsilon R_{v}, 
\end{align}
where,

\begin{align}
	\label{eq:perturbation_terms}
&R_{u} \equiv -\left(V(x)+i W(x)\right) u(x,t),  \\ \nonumber
&R_{v} \equiv -\left(V(x)+i W(x)\right) v(x,t).
\end{align} 

In the absence of the perturbation effect, i.e. $\epsilon = 0$, the Lagrangian density associated with Eq.~\eqref{eq:coupled_NLSE_new} is,
\begin{align}
	\label{eq:LagrangiandDensity}
\begin{split}
{\mathcal{L}} &=  \frac{i}{2} \left[u^* \frac{\partial u}{\partial t} -u \frac{\partial u^*}{\partial t}  \right] - \frac{1}{2} \left|  \frac{\partial u }{\partial x}  \right|^2 +\frac{1}{2} \left| u \right|^4  \\ 
& + \frac{i}{2} \left[v^* \frac{\partial v}{\partial t} -v \frac{\partial v^*}{\partial t}  \right] - \frac{1}{2} \left|  \frac{\partial v }{\partial x}  \right|^2 +\frac{1}{2} \left| v \right|^4  \\ 
&+ g \left| u \right|^2 \left| v \right|^2.
\end{split}
\end{align}

The use of a perturbation technique in the variational method results in modifying the standard Euler–Lagrange method. To determine the equations of motion that govern the behavior of the variational parameters, we use the following modified Euler–Lagrange equation~\cite{18}

\begin{equation}
	\label{eq:modified_Euler_Lagrange_eq}
	\frac{\partial L}{\partial a_{j}}  - \frac{d}{dt} \left(\frac{\partial L}{\partial \dot{a}_{j}}  \right) = 2\ \text{Re} \biggl\{ \int_{-\infty}^{\infty}( R^*_{u} \frac{\partial u }{\partial a_{j}} + R^*_{v} \frac{\partial v }{\partial a_{j}} ) \;dx  \biggl\}.
\end{equation}

Here, $L = \int_{-\infty}^{\infty} dx \mathcal{L}$, where $\mathcal{L}$ denotes the Lagrangian density in Eq.~\eqref{eq:LagrangiandDensity}. The variable $a_{j}$ represents the variational parameters, where $\dot{a}_{j} \equiv da/dt $ and $\text{Re} \{\}$ denote the real part of the expression in brackets. We employ the following ansatz as the variational BB soliton solution to the coupled NLSEs, Eq.~\eqref{eq:coupled_NLSE_new},

\begin{align}
\label{eq:ansatz}
u\left(x,t\right) &= A \mathrm{sech}\left(\frac{x+\xi_{1}}{a}\right) \mathrm{exp} \biggl\{i\left[ \phi +c_{1} \left(x+\xi_{1}\right) \right. \biggl. \\ \biggl.  \nonumber & +b \left(x+\xi_{1}\right)^2 \left. \right]\biggl\}, \\ \nonumber
v\left(x,t\right) &= A \mathrm{sech}\left(\frac{x+\xi_{2}}{a}\right) \mathrm{exp} \biggl\{i\left[ \phi +c_{2} \left(x+\xi_{2}\right) \right. \biggl. \\ \biggl.  \nonumber & +b \left(x+\xi_{2}\right)^2 \left. \right]\biggl\}.
\end{align}
The variational parameters $A(t)$, $a(t)$, $\phi(t)$, and $b(t)$ describe the amplitude, width, phase, and chirp of the two components, respectively. The location and velocity of the two components are represented by $\xi_{1}(t)$, $\xi_{2}(t)$, $c_{1}(t)$, and $c_{2}(t)$, respectively. We may link the amplitude to the width and reduce the number of variational parameters by one when we use the normalization condition.

\begin{align}
\label{eq:normalization}
\int_{-\infty}^{\infty} dx |u(x,t)|^2 = \int_{-\infty}^{\infty} dx |v(x,t)|^2 = 2 A^2 a = N.
\end{align}

To obtain a system of ordinary differential equations (ODEs) that describe the evolution of the variational parameters in time, we substitute Eq.~\eqref{eq:ansatz} into the Lagrangian density, Eq.~\eqref{eq:LagrangiandDensity}, and integrate over space from $-\infty$ to $+\infty$. As a result, we obtain the Lagrangian as a function of the variational parameters as 

\begin{align}
\label{eq:Lagrangian}
\mathrm{L} &= -\frac{N}{3a^2}+\frac{N^2}{3a}-\frac{1}{3}N \pi^2 a^2 b^2-\frac{1}{2} N \left(c^2_{1}+ c^2_{2}\right) -\frac{g N^2}{a^2} \times \nonumber \\ \nonumber
& \mathrm{csch}^2\left(\frac{\xi_{1}-\xi_{2}}{a}\right) \left[a-\left(\xi_{1}-\xi_{2}\right)\mathrm{coth}\left(\frac{\xi_{1}-\xi_{2}}{a}\right)\right] \\  
& -\frac{1}{6} N \pi^2 a^2 \frac{\partial b}{\partial t} -N c_{1} \frac{\partial \xi_{1}}{\partial t}-N c_{2} \frac{\partial \xi_{2}}{\partial t} -2 N \frac{\partial \phi}{\partial t}.
\end{align} 

The next step in our analysis involves applying the modified Euler-Lagrange equation (Eq.~\eqref{eq:modified_Euler_Lagrange_eq}, using Eq.~\eqref{eq:Lagrangian} in addition to Eqs.~\eqref{eq:perturbation_terms}).  Most of these equations are lengthy; hence, we relegate the writing of the explicit system of equations of motion to the Appendix. We used the variational equations of motion in the appendix to plot the trajectories of the two components.
   
\begin{figure}[!htbp]
\centerline{\includegraphics[width=0.8\columnwidth]{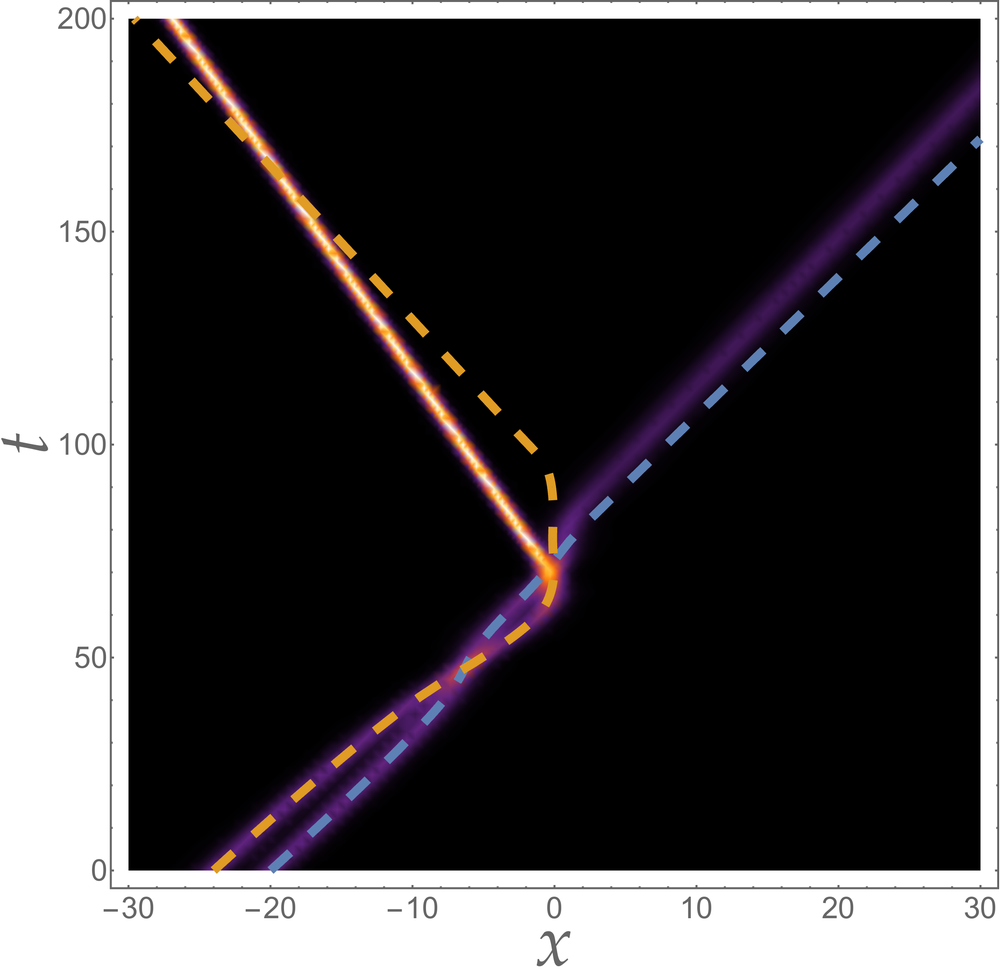}}
\caption{\label{fig:FRHPRA:figures/VA_num_1}  Density plot of a two-component BB soliton scattering by a $\mathcal{PT}$-symmetric potential, Eq.~\eqref{eq:real_imaginary_potential_2} with $\beta = -1$. The solid lines represent the numerical calculations for the two components, whereas the dashed lines represent the variational calculations. The potential parameters are: $V_{0} = -0.1$, $W_{0}=-0.08$. The locations of the two components are: $\xi_{1}=-20$, $\xi_{2}=-24$ with $v_{0} = 0.33$. The coupling strength, $g=0.05$.}
\end{figure}

\begin{figure}[!htbp]
\centerline{\includegraphics[width=0.8\columnwidth]{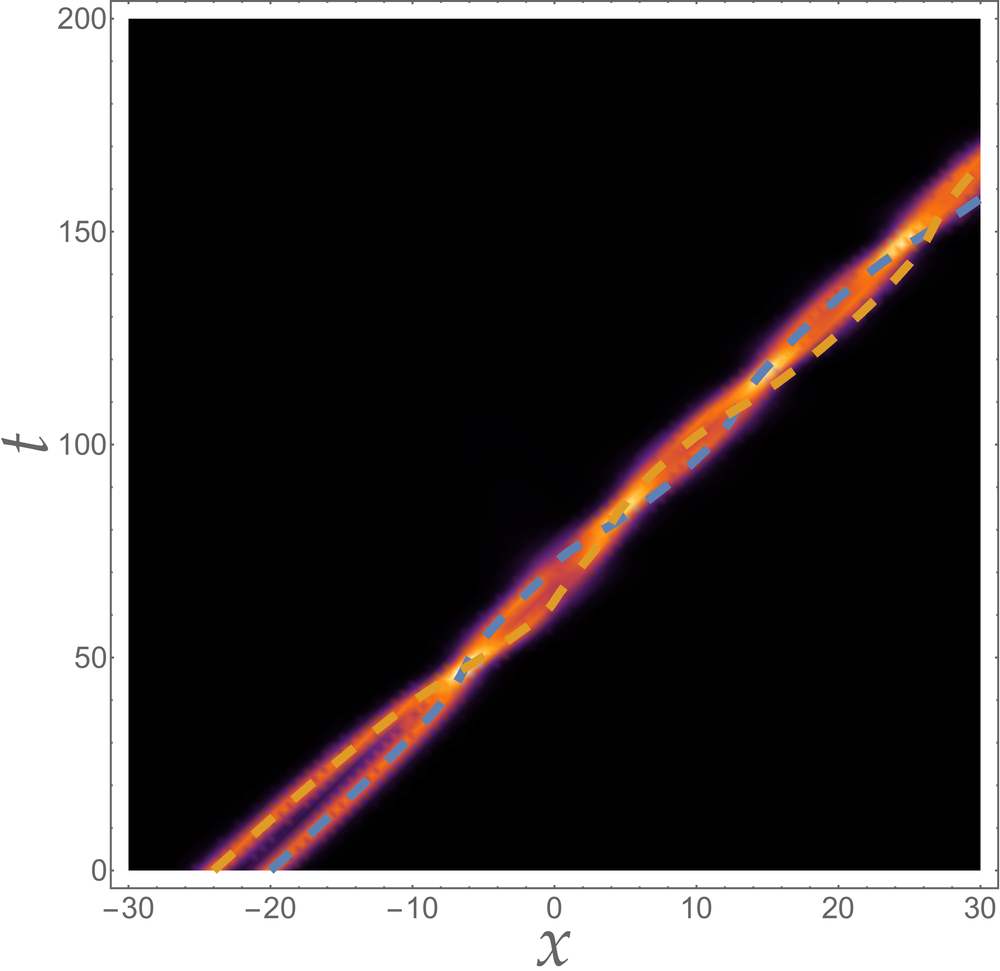}}
\caption{\label{fig:FRHPRA:figures/VA_num_2} Density plot of a two-component BB soliton scattering by a $\mathcal{PT}$-symmetric potential, Eq.~\eqref{eq:real_imaginary_potential_2} with $\beta = 1$. The solid lines represent the numerical calculations for the two components, whereas the dashed lines represent the variational calculations. The potential and BB soliton parameters are similar to Fig.~\ref{fig:FRHPRA:figures/VA_num_1}.}
\end{figure}

In Fig.~\ref{fig:FRHPRA:figures/VA_num_1} and~\ref{fig:FRHPRA:figures/VA_num_2}, the trajectory of the two components calculated from the variational calculations (dashed lines) were plotted and compared to the results of numerical simulation (solid lines). The two components underwent internal oscillation, where we set $g=0.05$. We obtain asymmetric dynamics, where unidirectional segregation is achieved using a delta function potential, Eq.~\eqref{eq:real_imaginary_potential_2}. Depending on whether the BB soliton first interacts with the gaining or damping term caused by the imaginary part of the $\mathcal{PT}$-symmetric potential, the two components break up or continue to oscillate. The two approaches are found to be in good agreement. In both figures, the two components' locations are $\xi_{1} = -20$ and $\xi_{2} = -24$ with $v_{0} = 0.33$. The potential parameters in Eq.~\eqref{eq:real_imaginary_potential_2} are $V_{0} = -0.1$, $W_{0} = -0.08$.

\begin{figure}[!h]
	\centering
\includegraphics[width=0.8\columnwidth]{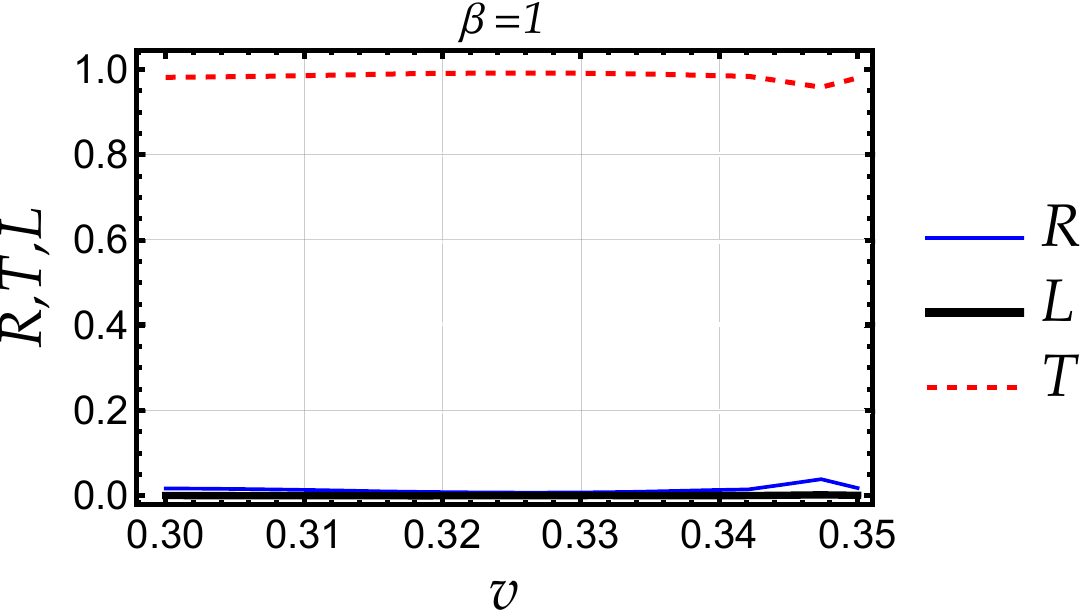}
\includegraphics[width=0.8\columnwidth]{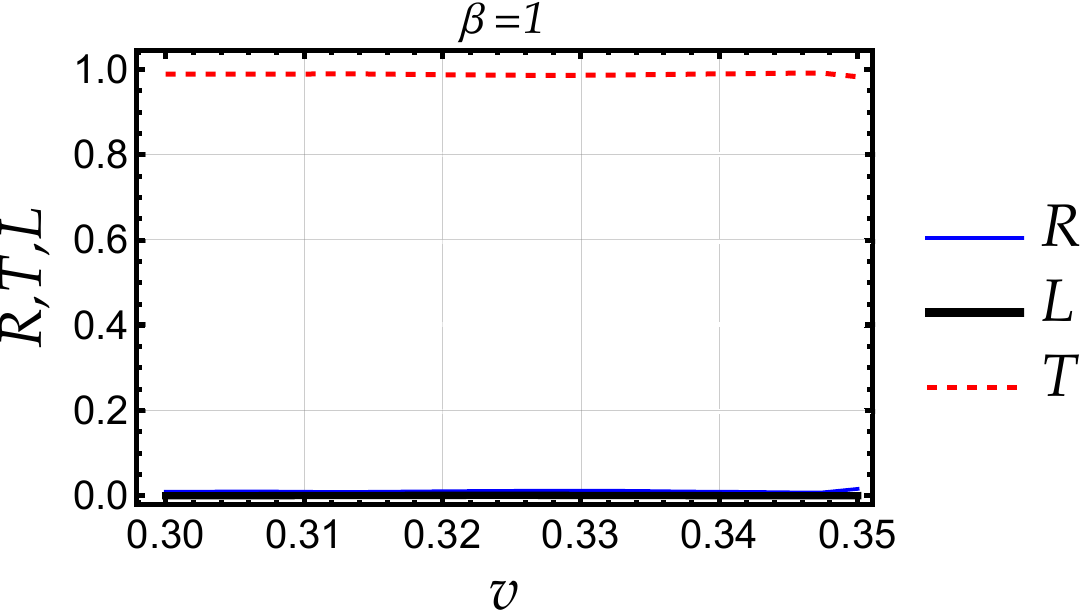}
	\caption{Transport coefficients obtained from the numerical solution of the coupled NLSEs, Eq.~\eqref{eq:coupled_NLSE}, using Eq.~\eqref{eq:real_imaginary_potential_2} with $\beta = 1$. The up (down) panel represents the first (second) component u(x,t) (v(x,t)). For the velocity range, both components transmitted over the potential.}
	\label{fig:FRHPRA:figures/TRL_num_delta_1_beta_pos_1}
\end{figure}

\begin{figure}[!h]
	\centering
\includegraphics[width=0.8\columnwidth]{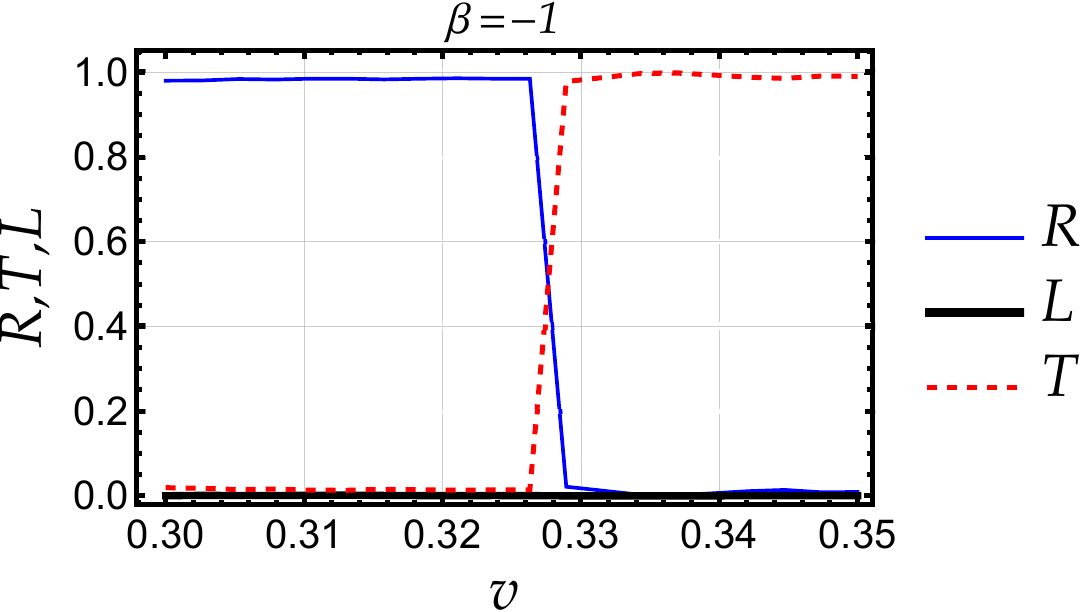}
\includegraphics[width=0.8\columnwidth]{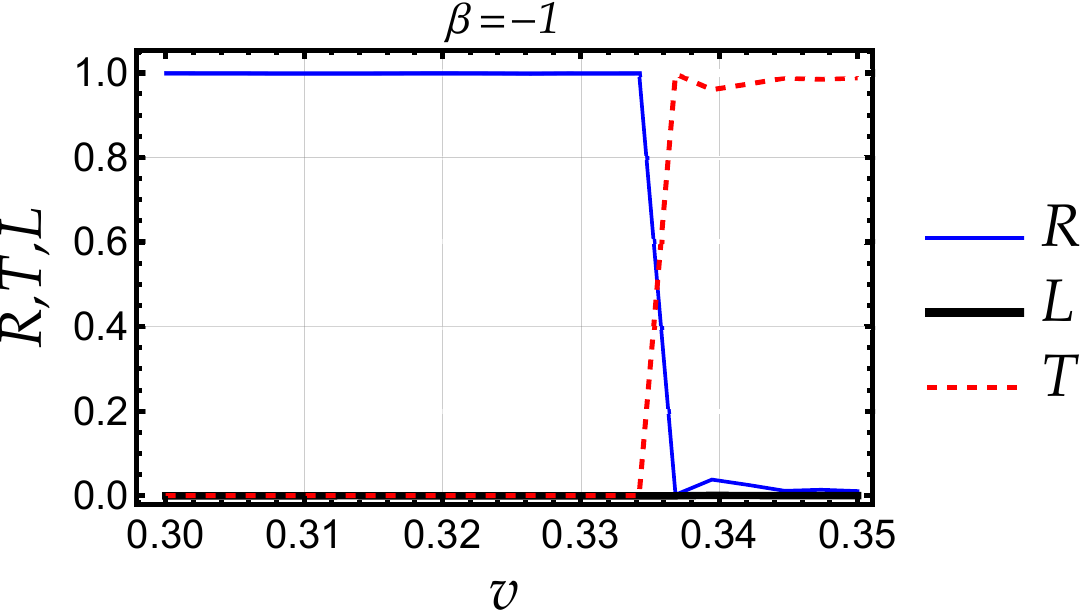}
	\caption{Transport coefficients obtained from the numerical solution of the coupled NLSEs, Eq.~\eqref{eq:coupled_NLSE}, using Eq.~\eqref{eq:real_imaginary_potential_2} with $\beta = -1$. The up (down) panel represents the first (second) component u(x,t) (v(x,t)). Depending on the incident velocity, we have a total reflection of both components, unidirectional segregation, or total transmission.}
	\label{fig:FRHPRA:figures/TRL_num_delta_1_beta_minus_1}
\end{figure}

Furthermore, we proceeded by expanding the comparison between the numerical simulation and variational calculations for a broad range of velocities. In Fig.~\ref{fig:FRHPRA:figures/TRL_num_delta_1_beta_pos_1}, we numerically calculated the transport coefficients for $\beta=1$ and found the two components transmitted over the potential. When we invert the potential around $x=0$ by setting $\beta=-1$, a small velocity window is achieved, in which unidirectional segregation of the two components is observed (see Fig. ~\ref{fig:FRHPRA:figures/TRL_num_delta_1_beta_minus_1}).

\begin{figure}[!h]
	\centering
\includegraphics[width=0.8\columnwidth]{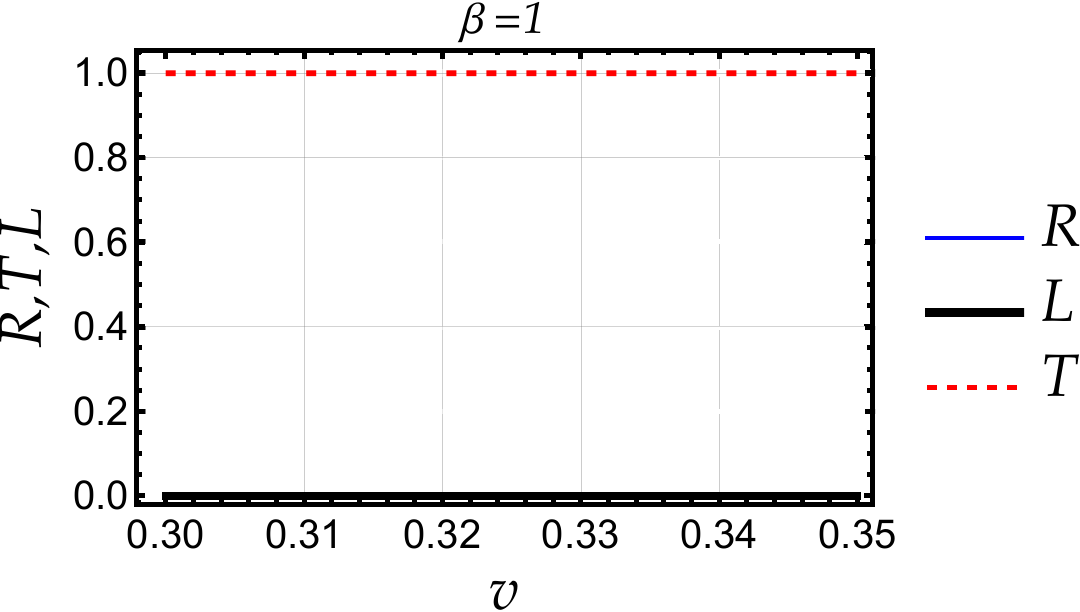}
\includegraphics[width=0.8\columnwidth]{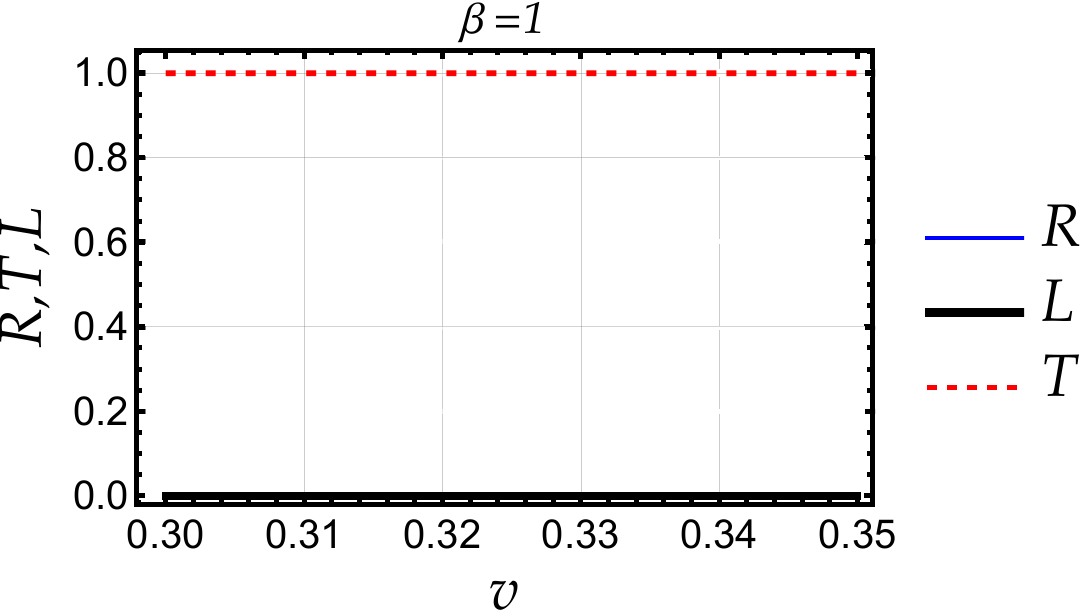}
	\caption{Transport coefficients obtained from the Variational calculations using Eq.~\eqref{eq:real_imaginary_potential_2} with $\beta = 1$. The up (down) panel represents the first (second) component u(x,t) (v(x,t)). For the velocity range, both components transmitted over the potential. }
	\label{fig:FRHPRA:figures/TRL_VA_delta_1_beta_pos_1}
\end{figure}

\begin{figure}[!h]
	\centering
\includegraphics[width=0.8\columnwidth]{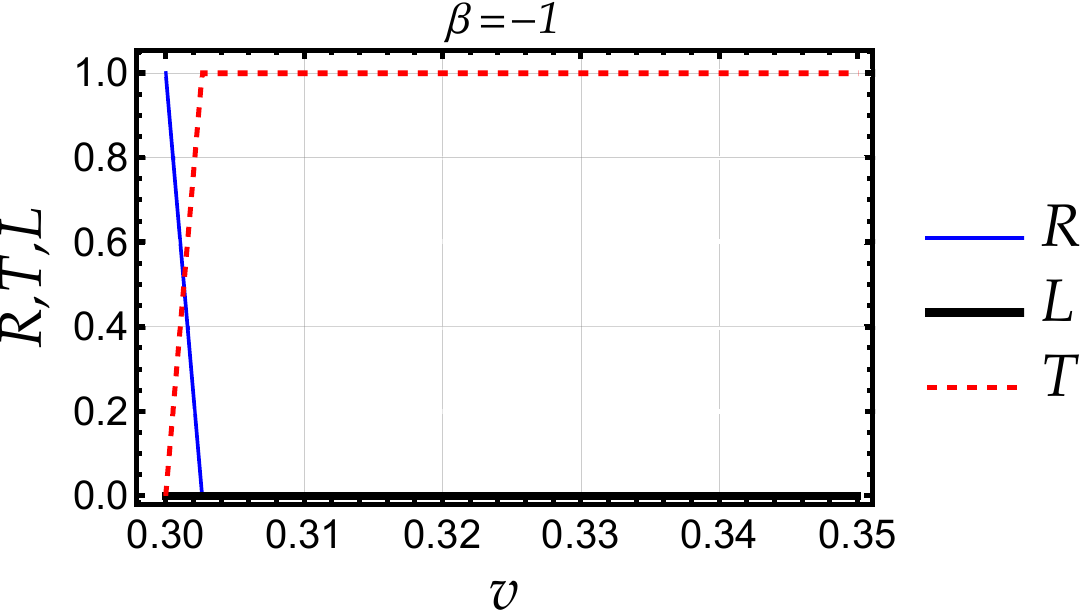}
\includegraphics[width=0.8\columnwidth]{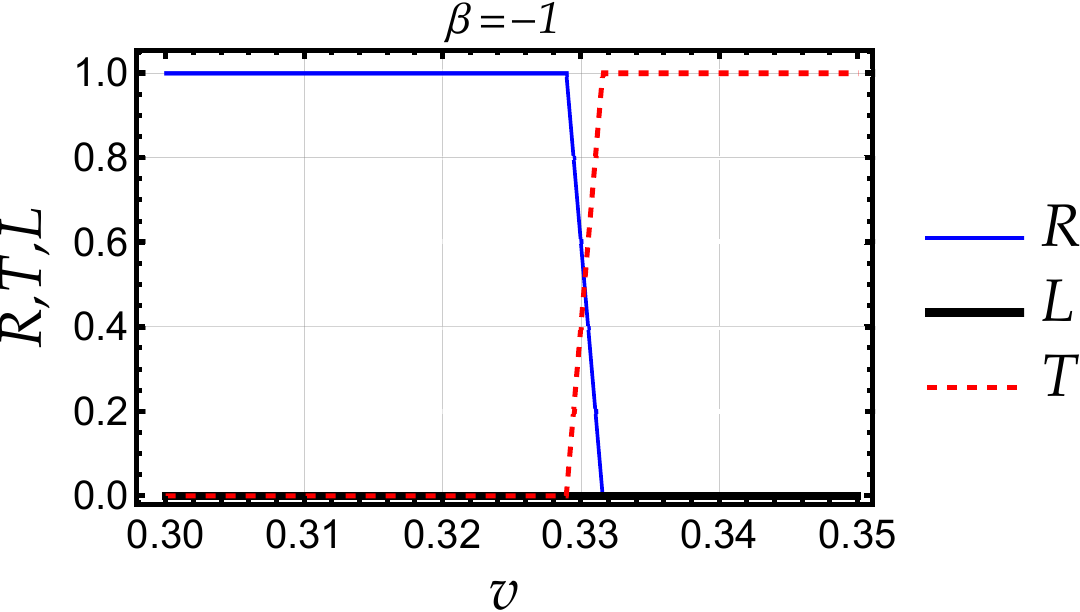}
	\caption{Transport coefficients obtained from the Variational calculations using Eq.~\eqref{eq:real_imaginary_potential_2} with $\beta = -1$. The up (down) panel represents the first (second) component u(x,t) (v(x,t)). Depending on the incident velocity, we have a total reflection of both components, unidirectional segregation, or total transmission.}
	\label{fig:FRHPRA:figures/TRL_VA_delta_1_beta_minus_1}
\end{figure}

In Fig.~\ref{fig:FRHPRA:figures/TRL_VA_delta_1_beta_pos_1} and~\ref{fig:FRHPRA:figures/TRL_VA_delta_1_beta_minus_1}, the transport coefficients were calculated analytically. Fig.~\ref{fig:FRHPRA:figures/TRL_VA_delta_1_beta_pos_1} shows that the same dynamics are obtained when setting $\beta=1$ (i.e., complete transmission of the two components over the potential). As shown in Fig. 10, we also obtain unidirectional segregation for a larger velocity window. A discrepancy between the numerical simulation and variational calculations is always expected, especially when using a sharp edge function, such as the delta function. The delta-function potential, Eq.~\eqref{eq:real_imaginary_potential_2}, captures the main $\mathcal{PT}$-symmetric features of the original potential, i.e., Eq.~\eqref{eq:real_imaginary_potential_1}.

\subsection*{Simplified dynamics}

The main aim of the variational calculation was to capture the physics of the unidirectional segregation of the BB soliton. To this end, we simplify the lengthy variational equations of motion in the appendix by focusing on the center-of-mass $\xi_{1,2}$ and the velocity $c_{1,2}$ of the two components. Consequently, we obtain the following coupled equations  
\
\begin{align}
	\label{eq:simplified_EOMs}
& \frac{\partial^2}{\partial t^2} \xi_{1}  +\Gamma\left[\xi_{1}\right] \frac{\partial}{\partial t} \xi_{1}  + f_{1}\left[\xi_{1},\xi_{2},c_{1}\right] =0,
\\ \nonumber
& \frac{\partial^2}{\partial t^2} \xi_{2}  +\Gamma\left[\xi_{2}\right] \frac{\partial}{\partial t} \xi_{2}  + f_{2}\left[\xi_{1},\xi_{2},c_{2}\right] =0, 
\end{align}
where,
\begin{align}
	\label{eq:Gammas_1}
	\nonumber
&\Gamma\left[\xi_{1,2}\right] = \epsilon \beta  \frac{W_{0}}{a}\left[\mathrm{sech}^2 \left(\frac{L-\xi_{1,2}}{a}\right) - \mathrm{sech}^2 \left(\frac{L+\xi_{1,2}}{a}\right) \right] \\ \nonumber 
&- \epsilon \beta  \frac{2W_{0}}{a^2} \left(L-\xi_{1,2}\right)\mathrm{sech}^2 \left(\frac{L-\xi_{1,2}}{a}\right) \mathrm{tanh} \left(\frac{L-\xi_{1,2}}{a}\right) \\ 
&+ \epsilon \beta  \frac{2W_{0}}{a^2} \left(L+\xi_{1,2}\right)\mathrm{sech}^2 \left(\frac{L+\xi_{1,2}}{a}\right) \mathrm{tanh} \left(\frac{L+\xi_{1,2}}{a}\right), 
\end{align}

and 
\begin{align}
	\label{eq:force_1}
	\nonumber
& f_{1}\left[\xi_{1},\xi_{2},c_{1}\right] = -\frac{3 g N}{a^2} \mathrm{coth} \left(\frac{\xi_{1}-\xi_{2}}{a}\right) \mathrm{csch}^2 \left(\frac{\xi_{1}-\xi_{2}}{a}\right) \\ \nonumber  
&+ \frac{g N}{a^3} \left(\xi_{1}-\xi_{2}\right) \left[2+ \mathrm{cosh} \left(2\frac{\xi_{1}-\xi_{2}}{a}\right) \right]  \mathrm{csch}^4  \left(\frac{\xi_{1}-\xi_{2}}{a}\right) \\ \nonumber 
&+ \frac{\epsilon \beta W_{0}}{a} \left\{\left[c_{1}+2 b \left(\xi_{1}-L\right)\right]\mathrm{sech}^2\left(\frac{L-\xi_{1}}{a}\right) \right.\\ \nonumber 
& \left. - \left[c_{1}+2 b \left(\xi_{1}+L\right)\right]\mathrm{sech}^2\left(\frac{L+\xi_{1}}{a}\right)\right\} \\ 
&+ \frac{\epsilon V_{0}}{a^2} \mathrm{sech}^2\left(\frac{\xi_{1}}{a}\right) \mathrm{tanh}\left(\frac{\xi_{1}}{a}\right), 
\end{align}
in addition to, 
\begin{align}
	\label{eq:force_2}
	\nonumber
& f_{2}\left[\xi_{1},\xi_{2},c_{2}\right] = \frac{3 g N}{a^2} \mathrm{coth} \left(\frac{\xi_{1}-\xi_{2}}{a}\right) \mathrm{csch}^2 \left(\frac{\xi_{1}-\xi_{2}}{a}\right) \\ \nonumber  
&- \frac{g N}{a^3} \left(\xi_{1}-\xi_{2}\right) \left[2+ \mathrm{cosh} \left(2\frac{\xi_{1}-\xi_{2}}{a}\right) \right]  \mathrm{csch}^4  \left(\frac{\xi_{1}-\xi_{2}}{a}\right) \\ \nonumber 
&+ \frac{\epsilon \beta W_{0}}{a} \left\{\left[c_{2}+2 b \left(\xi_{2}-L\right)\right]\mathrm{sech}^2\left(\frac{L-\xi_{2}}{a}\right) \right.\\ \nonumber 
& \left. - \left[c_{2}+2 b \left(\xi_{2}+L\right)\right]\mathrm{sech}^2\left(\frac{L+\xi_{2}}{a}\right)\right\} \\ 
&+ \frac{\epsilon V_{0}}{a^2} \mathrm{sech}^2\left(\frac{\xi_{2}}{a}\right) \mathrm{tanh}\left(\frac{\xi_{2}}{a}\right). 
\end{align}

The center-of-mass equations in Eqs.~\eqref{eq:simplified_EOMs} represent two classical particles subject to a velocity-dependent force $\Gamma[\xi_{1,2}]$, in addition to an effective force $-f_{1,2}\left[\xi_{1},\xi_{2},c_{1,2}\right]$. When the effective force is positive, the soliton velocity increases and is transmitted over the potential~\cite{11}. However, when the effective force is negative, the soliton velocity decreases and is consequently reflected by the barrier. The effective force depends on the location and velocity of the two components. Therefore, we can study the influence of one component on the transmission or reflection of the other component. Furthermore, by following the same approach in this study, where we fixed the launching point of the BB soliton to the left of the $\mathcal{PT}$-symmetric potential, we can plot the effective force in this region as a function of the soliton position.

\begin{figure}[!htbp]
\centerline{\includegraphics[width=\columnwidth]{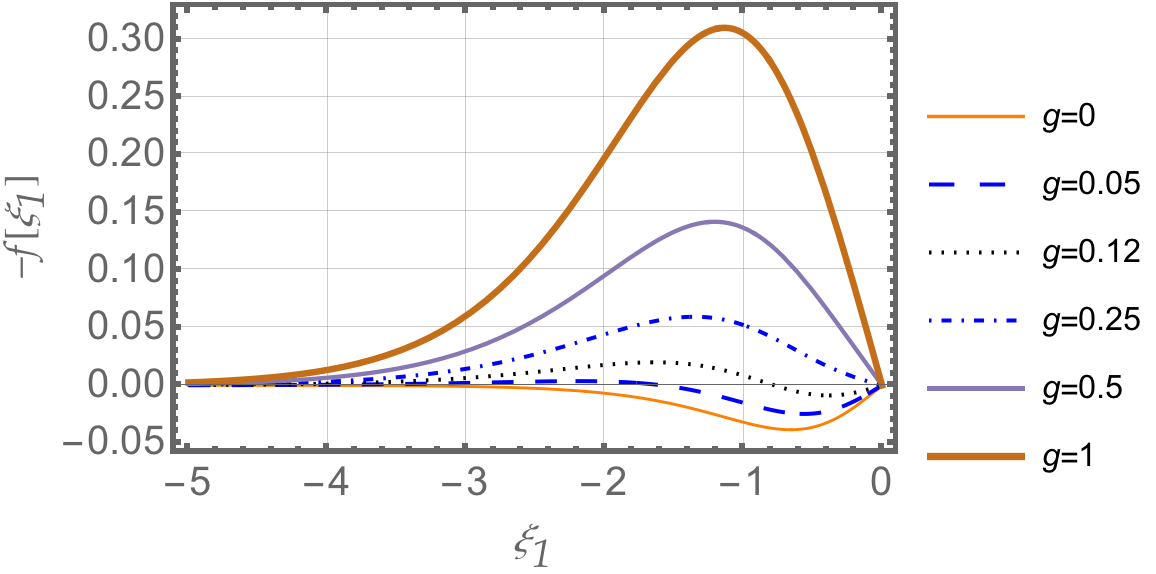}} 
\caption{\label{fig:FRHPRA:figures/force_3}  Effective force, $-f_{1} \left[\xi_{1}\right]$, for different coupling strength, $g$. Parameters are as follows: $V_{0}=-0.1$, $W_{0}=-0.08$, $a=1$, $\xi_{2}=0$, $N=2$, $c_{1}=0$, and $L=0.5$. } 
\end{figure}

In Fig.~\ref{fig:FRHPRA:figures/force_3}, we fixed one of the solitons at the center, $x = 0$, and plotted the effective force versus $\xi_{1}$ for different values of coupling strength. For the uncoupled case (i.e., $g=0$), the effective force is negative, which causes the soliton to slow down and reflect. By setting $g=0.05$, the case in Fig.~\ref{fig:FRHPRA:figures/VA_num_1}, we observe that the soliton is also reflected because the effective force is negative. Nevertheless, as we increase the coupling strength, the effective force changes sign from negative to positive, which means that the soliton in this case experiences an increase in velocity. Consequently, the soliton was transmitted. The splitting of a strong coupling BB soliton is difficult owing to the attractive interaction between the two components. Nonetheless, here we found that this behavior can be explained by considering the signs of the effective forces. Consequently, as shown in Fig.~\ref{fig:FRHPRA:figures/force_3}, we find that it is possible to split the two components as long as the coupling strength does not exceed $g\approx 0.12$

\section{Conclusions}
\label{sec:Conclusions}
In this study, the unidirectional splitting of a moving BB soliton through $\mathcal{PT}$-symmetric potentials was numerically demonstrated. Using an odd function in the imaginary part of the $\mathcal{PT}$-symmetric potential causes asymmetry in the ordering of pumping and damping. A BB soliton coming from one direction may experience a velocity increase or decrease depending on whether it encounters the pumping or damping region. We found that using reflection-less type $\mathcal{PT}$-symmetric potentials results in a sharp transition in the transport coefficients, allowing control of the transmission or reflection of each BB soliton component separately, thereby achieving unidirectional segregation. In addition, considering a $\mathcal{PT}$-symmetric potential with a delta function in the real and imaginary parts allows us to treat the potential as a small perturbation. Therefore, we can use a modified perturbed dynamical variational Lagrangian approximation to obtain the equations of motion of the system. The results from the variational approach were in good agreement with the numerical calculations. A simplified variational equation of motion shows that the two soliton components behave as classical particles subject to velocity-dependent and effective forces. By examining the effective force, we determined the upper limit of the coupling strength constant $g$, which enabled the splitting of the two components in the BB soliton. Future work may extend our study to examine unidirectional segregation of BB solitons using nonlinear management.

\section*{Acknowledgment}
The author thanks U. Al-Khawaja for the useful discussions.

\appendix
\section{Variational equations of motion}
%
\label{appendix} 

By using the Lagrangian equation, Eq.(\ref{eq:Lagrangian}), the modified Euler-Lagrange equations yield the following variational equations
of motion. We use this system of equations to plot the trajectory of the two components.
\begin{align}
    \label{EOMS}
\begin{split}
& \frac{2 N}{3 a^3}-\frac{N^2}{3 a^2}-\frac{2}{3}N \pi^2 a b^2-\frac{1}{3} N \pi^2 a \frac{\partial b}{\partial t} +\frac{g N^2}{a^2} \mathrm{csch}\left(\frac{\xi_{1}-\xi_{2}}{a}\right) \\ \nonumber & 
- \frac{4 g N^2}{a^3} \left(\xi_{1}-\xi_{2}\right) \mathrm{coth}\left(\frac{\xi_{1}-\xi_{2}}{a}\right) \mathrm{csch}^2\left(\frac{\xi_{1}-\xi_{2}}{a}\right) \\ \nonumber &  
- \frac{\epsilon N V_{0}}{2 a^3} \left\{ a\left[\mathrm{sech}^2\left(\frac{\xi_{1}}{a}\right)+\mathrm{sech}^2\left(\frac{\xi_{2}}{a}\right)\right] -2 \left[ \xi_{1} \mathrm{sech}^2\left(\frac{\xi_{1}}{a}\right) \right. \right. \nonumber \\  & \left.  \left.  \times \mathrm{tanh}\left(\frac{\xi_{1}}{a}\right) +\xi_{2} \mathrm{sech}^2\left(\frac{\xi_{2}}{a}\right) \mathrm{tanh}\left(\frac{\xi_{2}}{a}\right) \right] \right\} +\frac{g N^2}{a^4} \\ &
\times \left(\xi_{1}-\xi_{2}\right)^2 \mathrm{csch}^4 \left(\frac{\xi_{1}-\xi_{2}}{a}\right) \left[2+\mathrm{cosh}\left(2\frac{\xi_{1}-\xi_{2}}{a}\right)\right] = 0,
\end{split}
\\
\\
\begin{split}
& \frac{3 g N^2}{a^2} \mathrm{coth} \left(\frac{\xi_{1}-\xi_{2}}{a}\right) \mathrm{csch}^2 \left(\frac{\xi_{1}-\xi_{2}}{a}\right) +\epsilon \left\{
-\frac{N V_{0}}{a^2} \right. \\  & \left. \times \mathrm{sech}^2 \left(\frac{\xi_{1}}{a}\right)\mathrm{tanh} \left(\frac{\xi_{1}}{a}\right)
+ \frac{N W_{0} \beta}{a} \left[\mathrm{sech}^2\left(\frac{L-\xi_{1}}{a}\right) \right. \right.  \\  & \left.  \left.  \times
 \left[-c_{1}+2b\left(L-\xi_{1}\right)\right]  + \mathrm{sech}^2\left(\frac{L+\xi_{1}}{a}\right)  
 \left[c_{1}+2b\left(L+\xi_{1}\right)\right]\right]\right\} \\  &
 -\frac{2 g N^2}{a^3} \mathrm{coth}^2 \left(\frac{\xi_{1}-\xi_{2}}{a}\right) \mathrm{csch}^2  \left(\frac{\xi_{1}-\xi_{2}}{a}\right)  \left(\xi_{1}-\xi_{2}\right) \\  &
- \frac{g N^2}{a^3} \mathrm{csch}^4 \left(\frac{\xi_{1}-\xi_{2}}{a}\right) \left(\xi_{1}-\xi_{2}\right) + N \frac{\partial c_{1}}{\partial t} = 0,
\end{split}
\\
\begin{split}
& -\frac{3 g N^2}{a^2} \mathrm{coth} \left(\frac{\xi_{1}-\xi_{2}}{a}\right) \mathrm{csch}^2 \left(\frac{\xi_{1}-\xi_{2}}{a}\right) +\epsilon \left\{
-\frac{N V_{0}}{a^2} \right. \\  & \left. \times \mathrm{sech}^2 \left(\frac{\xi_{2}}{a}\right)\mathrm{tanh} \left(\frac{\xi_{2}}{a}\right)
+ \frac{N W_{0} \beta}{a} \left[\mathrm{sech}^2\left(\frac{L-\xi_{2}}{a}\right) \right. \right.  \\  & \left.  \left.  \times
 \left[-c_{2}+2b\left(L-\xi_{2}\right)\right]  + \mathrm{sech}^2\left(\frac{L+\xi_{2}}{a}\right)  
 \left[c_{2}+2b\left(L+\xi_{2}\right)\right]\right]\right\} \\  &
 +\frac{2 g N^2}{a^3} \mathrm{coth}^2 \left(\frac{\xi_{1}-\xi_{2}}{a}\right) \mathrm{csch}^2  \left(\frac{\xi_{1}-\xi_{2}}{a}\right)  \left(\xi_{1}-\xi_{2}\right) \\  &
+ \frac{g N^2}{a^3} \mathrm{csch}^4 \left(\frac{\xi_{1}-\xi_{2}}{a}\right) \left(\xi_{1}-\xi_{2}\right) + N \frac{\partial c_{2}}{\partial t} = 0,
\end{split}
\\
\begin{split}
& \frac{\partial \xi_{1}}{\partial t} + c_{1} -\frac{\epsilon W_{0} \beta}{a} \left[\left(L-\xi_{1}\right)\mathrm{sech}^2\left(\frac{L-\xi_{1}}{a}\right)  \right.   \\  & 
+\left. \left(L+\xi_{1}\right) \mathrm{sech}^2\left(\frac{L+\xi_{1}}{a}\right)\right] = 0,
\end{split}
\\
\begin{split}
& \frac{\partial \xi_{2}}{\partial t} + c_{2} -\frac{\epsilon W_{0} \beta}{a} \left[\left(L-\xi_{2}\right)\mathrm{sech}^2\left(\frac{L-\xi_{2}}{a}\right)
\right.   \\  &  \left.
+\left(L+\xi_{2}\right) \mathrm{sech}^2\left(\frac{L+\xi_{2}}{a}\right)\right] = 0,
\end{split}
\\
\begin{split}
& -\frac{2}{3}N \pi^2 a^2 b+ \frac{\epsilon N W_{0} \beta}{a} \left[ -\mathrm{sech}^2\left(\frac{L-\xi_{1}}{a}\right) \left(L-\xi_{1}\right)^2 
\right.  \\  &   \left. + \mathrm{sech}^2\left(\frac{L+\xi_{1}}{a}\right) \left(L+\xi_{1}\right)^2  -\mathrm{sech}^2\left(\frac{L-\xi_{2}}{a}\right) \left(L-\xi_{2}\right)^2 \right. \\  &   \left.
+ \mathrm{sech}^2\left(\frac{L+\xi_{2}}{a}\right) \left(L+\xi_{2}\right)^2\right] + \frac{1}{3} N \pi^2 a \frac{\partial a}{\partial t}=0 
\end{split}
\end{align}
\newpage 

\end{document}